\newcommand{\DoPrePrint}{0} 
\documentclass[%
reprint,
superscriptaddress,
showpacs,
preprintnumbers,
showkeys,
nofootinbib,
 amsmath,amssymb,
 aps,
 prd,
floatfix,
nofootinbib,
]{revtex4-1}

\usepackage{xspace}
\usepackage{graphicx}% Include figure files
\usepackage{array}
\usepackage{dcolumn}% Align table columns on decimal point
\usepackage{bm}% bold math
\usepackage{rotating}
\usepackage{multirow}
\usepackage{xcolor}

\usepackage{hyperref}% add hypertext capabilities 

\usepackage[mathlines]{lineno}% Enable numbering of text and display math
\ifnum\DoPrePrint=1
\linenumbers\relax % Commence numbering lines
\fi

\newcommand\minerva  {MINERvA\xspace}

\begin{document}

\title{Data Preservation at \minerva} 

%% List of institution addresses, in command form.
\newcommand{\Rutgers}{Rutgers, The State University of New Jersey, Piscataway, New Jersey 08854, USA}
\newcommand{\Hampton}{Hampton University, Dept. of Physics, Hampton, VA 23668, USA}
\newcommand{\Dortmund}{Institute of Physics, Dortmund University, 44221, Germany }
\newcommand{\Otterbein}{Department of Physics, Otterbein University, 1 South Grove Street, Westerville, OH, 43081 USA}
\newcommand{\JMU}{James Madison University, Harrisonburg, Virginia 22807, USA}
\newcommand{\Florida}{University of Florida, Department of Physics, Gainesville, FL 32611}
\newcommand{\UCIrvine}{Department of Physics and Astronomy, University of California, Irvine, Irvine, California 92697-4575, USA}
\newcommand{\CBPF}{Centro Brasileiro de Pesquisas F\'{i}sicas, Rua Dr. Xavier Sigaud 150, Urca, Rio de Janeiro, Rio de Janeiro, 22290-180, Brazil}
\newcommand{\PUCP}{Secci\'{o}n F\'{i}sica, Departamento de Ciencias, Pontificia Universidad Cat\'{o}lica del Per\'{u}, Apartado 1761, Lima, Per\'{u}}
\newcommand{\INRM}{Institute for Nuclear Research of the Russian Academy of Sciences, 117312 Moscow, Russia}
\newcommand{\Jlab}{Jefferson Lab, 12000 Jefferson Avenue, Newport News, VA 23606, USA}
\newcommand{\Pittsburgh}{Department of Physics and Astronomy, University of Pittsburgh, Pittsburgh, Pennsylvania 15260, USA}
\newcommand{\Guanajuato}{Campus Le\'{o}n y Campus Guanajuato, Universidad de Guanajuato, Lascurain de Retana No. 5, Colonia Centro, Guanajuato 36000, Guanajuato M\'{e}xico.}
\newcommand{\Athens}{Department of Physics, University of Athens, GR-15771 Athens, Greece}
\newcommand{\Tufts}{Physics Department, Tufts University, Medford, Massachusetts 02155, USA}
\newcommand{\WM}{Department of Physics, William \& Mary, Williamsburg, Virginia 23187, USA}
\newcommand{\FNAL}{Fermi National Accelerator Laboratory, Batavia, Illinois 60510, USA}
\newcommand{\Purdue}{Department of Chemistry and Physics, Purdue University Calumet, Hammond, Indiana 46323, USA}
\newcommand{\MCLA}{Massachusetts College of Liberal Arts, 375 Church Street, North Adams, MA 01247}
\newcommand{\UMD}{Department of Physics, University of Minnesota -- Duluth, Duluth, Minnesota 55812, USA}
\newcommand{\Northwestern}{Northwestern University, Evanston, Illinois 60208}
\newcommand{\UNI}{Facultad de Ciencias, Universidad Nacional de Ingenier\'{i}a, Apartado 31139, Lima, Per\'{u}}
\newcommand{\Rochester}{Department of Physics and Astronomy, University of Rochester, Rochester, New York 14627 USA}
\newcommand{\Austin}{Department of Physics, University of Texas, 1 University Station, Austin, Texas 78712, USA}
\newcommand{\USM}{Departamento de F\'{i}sica, Universidad T\'{e}cnica Federico Santa Mar\'{i}a, Avenida Espa\~{n}a 1680 Casilla 110-V, Valpara\'{i}so, Chile}
\newcommand{\Geneva}{University of Geneva, 1211 Geneva 4, Switzerland}
\newcommand{\Chicago}{Enrico Fermi Institute, University of Chicago, Chicago, IL 60637 USA}
\newcommand{\hired}{}
\newcommand{\OregonState}{Department of Physics, Oregon State University, Corvallis, Oregon 97331, USA}
\newcommand{\oxford}{Oxford University, Department of Physics, Oxford, OX1 3PJ United Kingdom}
\newcommand{\umiss}{University of Mississippi, Oxford, Mississippi 38677, USA}
\newcommand{\upenn}{Department of Physics and Astronomy, University of Pennsylvania, Philadelphia, PA 19104}
\newcommand{\AMU}{AMU Campus, Aligarh, Uttar Pradesh 202001, India}
\newcommand{\wroclaw}{University of Wroclaw, plac Uniwersytecki 1, 50-137 Wroa\l{}aw, Poland}
\newcommand{\Mohali}{Department of Physical Sciences, IISER Mohali, Knowledge City, SAS Nagar, Mohali - 140306, Punjab, India}
\newcommand{\york}{York University, Department of Physics and Astronomy, Toronto, Ontario, M3J 1P3 Canada}
\newcommand{\ND}{Department of Physics, University of Notre Dame, Notre Dame, Indiana 46556, USA}
\newcommand{\ICL}{The Blackett Laboratory,  Imperial College London,  London SW7 2BW, United Kingdom}
\newcommand{\warwick}{Department of Physics, University of Warwick, Coventry, CV4 7AL, UK}

\newcommand{\finerThanks}{Now at Los Alamos National Laboratory, Los Alamos, New Mexico 87545, USA}
\newcommand{\bamThanks}{Now at University of Minnesota, Minneapolis, Minnesota 55455, USA}
\newcommand{\mascencioThanks}{Now at Iowa State University, Ames, IA 50011, USA}
\newcommand{\kleykampThanks}{now at Department of Physics and Astronomy, University of Mississippi, Oxford, MS 38677}
\newcommand{\byaeggyThanks}{Now at Department of Physics, University of Cincinnati,  Cincinnati, Ohio 45221, USA}
\newcommand{\contactInfo}{Please direct correspondence to finer@fnal.gov, mess@umn.edu}

% 25 total signatories.

\author{R.~Fine}\thanks{\finerThanks}				\affiliation{\Rochester}
\author{B.~Messerly}\thanks{\bamThanks}			\affiliation{\Pittsburgh}
\author{K.S.~McFarland}							\affiliation{\Rochester}
\author{S.~Akhter}								\affiliation{\AMU}
\author{V.~Ansari}								\affiliation{\AMU}
\author{M.~V.~Ascencio}\thanks{\mascencioThanks}		\affiliation{\PUCP}
\author{J.~L.~Bonilla}							\affiliation{\Guanajuato}
\author{H.~da~Motta}							\affiliation{\CBPF}
\author{L.~Fields}								\affiliation{\ND}
\author{R.~Gran}								\affiliation{\UMD}
\author{E.Granados}								\affiliation{\Guanajuato}
\author{D.A.~Harris}								\affiliation{\york}  \affiliation{\FNAL}
\author{D.~Jena}								\affiliation{\FNAL}
\author{S.~Jena}								\affiliation{\Mohali}
\author{J.~Kleykamp}\thanks{\kleykampThanks}		\affiliation{\Rochester}
\author{A.~Klustov\'{a}}							\affiliation{\ICL}
\author{X.-G.~Lu}								\affiliation{\warwick}  \affiliation{\oxford}
\author{W.A.~Mann}								\affiliation{\Tufts}
\author{A.~Olivier}								\affiliation{\Rochester}
\author{K.-J.~Plows}								\affiliation{\oxford}
\author{D.~Ruterbories}							\affiliation{\Rochester}
\author{H.~Schellman}							\affiliation{\OregonState}
\author{H.~Su}									\affiliation{\Pittsburgh}
\author{E.~Valencia}								\affiliation{\WM}  \affiliation{\Guanajuato}
\author{B.~Yaeggy}\thanks{\byaeggyThanks}			\affiliation{\USM}

\collaboration{The MINERvA Collaboration} \email{\contactInfo}

\date{September 13, 2022}

\begin{abstract}

Between 2013 and 2019 \minerva collected an accelerator neutrino interaction dataset that is uniquely relevant to the energy range of DUNE. These are the only currently available data at intermediate and high momentum transfers for multiple nuclear targets in the same beam. \minerva is undertaking a campaign to preserve these data and make them publicly available so that they may be analyzed beyond the end of the \minerva collaboration. We encourage the community to consider the development of centralized resources to enable long-term access to these data and analysis tools for the entire HEP community. 

\end{abstract}

\maketitle

\section{Technical Strategy} \label{technicalStrategy}

The \minerva data preservation project consists of three components: (1) preservation of \minerva data into a single ROOT tuple that incorporates low- and high-level reconstructed objects; (2) the MINERvA Analysis Toolkit (MAT) -- a broadly applicable HEP software toolkit for calculating systematic uncertainties using event tuple objects; (3) a software package built on the MAT for reproducing \minerva published results, which includes templates for performing new analyses. 

To date, each published \minerva analysis has employed its own tailored strategy for preparing ROOT tuples that feed into the ``macro'' (event-loop) stage of analysis. They are prepared using the Gaudi framework \cite{Barrand:2001ny}, and apply analysis-specific reconstruction techniques to commonly calibrated and prepared low-level data. Historically, this has served \minerva well, as it enables parallel development of distinct reconstruction techniques and decentralizes the production of analysis tuples. The \minerva analysis program has reached a level of maturity at which we can now summarize the reconstruction for a broad variety of final states in a unified analysis tuple. This structure will support a large number of analyses using a smaller disk footprint than would be used by separately preserving tuples which are specialized to individual analyses. This approach also obviates the need to develop new Gaudi routines, which further reduces the computational resources needed to perform future analyses. We will include in these tuples low-level reconstruction objects that could, in principle, be used for novel reconstructions.

It is a common analysis strategy in HEP experiments to estimate systematic uncertainties by simulating the experiment many times in a multitude of systematic ``universes''. In \minerva analyses, systematic universes incorporate the effects of systematic uncertainties by inserting variations directly into physics distributions at all stages of the analyses. For example, a typical analysis at MINERvA combines the distributions of signal and background events as functions of some kinematic variable, $S(x), B(x)$, the efficiency of performing that selection, $\epsilon(x)$, and the integrated flux, $\Phi$, to extract a differential cross section, $\frac{d\sigma}{dx} \propto \frac{S(x)-B(x)}{\epsilon(x)\Phi}$. Each of these inputs is calculated independently in $\mathcal{O}(100)$ distinct systematic universes, and is stored in a modified version of ROOT's TH\{1,2\}D class. Using these objects, it is straightforward to calculate the uncertainty arising from any subset of the systematic variations for any of these physics distributions, or any distribution derived from them. Similarly, any analysis technique used in the extraction of a cross section, such as the construction of background sideband constraints, is performed independently in each systematic universe. A suite of custom C++ classes facilitates the execution of this strategy, and streamlines the evaluation of systematic uncertainty across all MINERvA analyses. Collectively, we refer to these classes as the MINERvA Analysis Toolkit (MAT). As part of our data preservation effort, we intend to make the MAT publicly available and we encourage its adoption by other neutrino experiments.

Using the methods provided by the MAT, writing event loops is straightforward. Within the loop over a ROOT tuple, there is a loop over systematic universes. In each systematic universe, cuts are applied and histograms corresponding to various kinematic variables are filled. This preserves the event-by-event effects of the systematic variations across all bins of each histogram constructed. Whereas a systematics-agnostic user would fill a TH1D, for example, in this event loop, the user instead fills an ``MnvH1D'', which maps a TH1D to each systematic variation. Downstream, the MnvH1D supports the standard operations available for a TH1D and executes them across all systematic universes. In general, a systematic variation may modify the value of a kinematic variable, and alter an event selection through a cut placed on that kinematic variable. By incorporating the handling of systematic universes into the event loop, this approach guarantees that kinematic variables are shifted appropriately and that the effects of those shifts are propagated to all downstream aspects of the analysis. As part of our data preservation campaign, we are developing software using this approach that will easily reproduce \minerva analyses and enable future modifications to them. Future use-cases may include the modification of an event selection, the construction of a new observable, or the implementation of a new signal definition. Future users will be able to make any of these modifications and re-extract cross sections. For example, a future user may be interested in adding a final state neutron requirement to $\nu_{\mu}$ CCQE-like final states, defining the transverse angle between that neutron's direction and the summed $p_{T}$ of the leading proton and muon, and then extracting a differential cross-section with respect to that new variable.

\section{Applications of MINERvA Data}

\minerva completed data-taking in 2019, and expects the number of analyses undertaken by the collaboration to dramatically curtail soon. Once the DUNE near detector begins to collect data (later in this decade \cite{Acciarri:2016crz}), its data set (FHC) will likely exceed the size of MINERvA's within one year of operations \cite{Marshall:2019vdy}. Thus, there will be at least five years during which the \minerva data set provides the community's only opportunity to study neutrino interactions at intermediate and high momentum transfers for multiple nuclear targets in the same beam. The \minerva collaboration is likely to continue in a less active configuration during this interim period, but recognizes that it will have insufficient resources available to address all questions that may arise. Therefore, we believe it is important to maintain access to the data \minerva has collected to continue probing the interaction models that will be used by DUNE in the measurement of CP violation and other neutrino phenomena. 

In recent years, \minerva has led the field in probing these models in the few-GeV regime. To date we have published more than 30 cross section and flux measurements using our ``Low Energy'' (LE; $\langle E_{\nu}\rangle\sim 3$ GeV) data \cite{MINERvA:2020zzv,MINERvA:2020anu,MINERvA:2019ope,MINERvA:2019rhx,MINERvA:2019wqe,MINERvA:2018hqn,MINERvA:2018hba,MINERvA:2018nab,MINERvA:2018vjb,MINERvA:2017ipy,MINERvA:2017okh,MINERvA:2017dzh,MINERvA:2017ozn,MINERvA:2016ymg,MINERvA:2016ing,MINERvA:2016cun,MINERvA:2016sfc,MINERvA:2016zyp,MINERvA:2016uck,MINERvA:2016oql,MINERvA:2015nqi,MINERvA:2015ydy,MINERvA:2015jih,MINERvA:2015slz,MINERvA:2014ypj,MINERvA:2014ani,MINERvA:2014ogb,MINERvA:2014rdw,MINERvA:2013bcy,MINERvA:2013kdn}, and we expect that our ``Medium Energy'' (ME; $\langle E_{\nu}\rangle\sim6$ GeV) data would support at least as many publications (a number of which are already complete \cite{MINERvA:2022mnw,MINERvA:2021wjs,MINERvA:2021dhf,MINERvA:2021owq,MINERvA:2019gsf,MINERvA:2019hhc}). Together, these data sets include over 4 million charged-current $\nu_{\mu}$ interactions in the active ($CH$) region of the detector, and roughly half as many $\bar{\nu}_{\mu}$ interactions. There are an additional $\mathcal{O}(1\text{ million})$ interactions in the passive nuclear target region of the detector, which includes $He$, $H_{2}O$, $C$, $Fe$, and $Pb$. These data can be analyzed to support the construction of interaction models that span nuclei both larger than and smaller than argon. Though we have used and continue to use these data, current efforts are limited by the human resources available to analyze them. 

In the coming years, neutrino interaction models must be improved to ensure the success of DUNE's ambitious physics program. Until then, \minerva will offer the largest and most relevant neutrino interaction dataset, against which such models can be tested. New, discriminating analysis techniques are continually being invented and refined, and the \minerva data have the flexibility to be used for studying new observables. For example, consider the analysis technique in which transverse kinematic imbalance is used to probe intranuclear dynamics in neutrino interactions~\cite{Lu:2015tcr, Furmanski:2016wqo, Abe:2018pwo, Dolan:2018sbb, Lu:2018stk, Dolan:2018zye, Lu:2019nmf, Harewood:2019rzy, Cai:2019jzk, Cai:2019hpx, Coplowe:2020yea} or the absence thereof~\cite{Lu:2015hea, Duyang:2018lpe, Duyang:2019prb, Munteanu:2019llq, Hamacher-Baumann:2020ogq}. This technique has provided a new handle on probing nuclear effects, but has only been utilized in the analysis of modern data sets. Evidently, the infrastructure is not available to re-analyze historical data at this level of detail. We believe that both access to the data and to an infrastructure to facilitate its analysis are necessary components of a successful preservation campaign. As described in Section \ref{technicalStrategy}, our data preservation strategy includes support for the calculation of new kinematic variables and access to software which will enable reproduction of a wide range of current \minerva analyses. We expect this to serve as a launching point for the reanalysis of existing selections to include new kinematic variables or to test against new interaction models. We also expect that future analyzers may modify existing selections to measure additional classes of neutrino interactions. For example, a future analyzer may wish to further constrain our $\nu_{\mu}$ CCQE-like selection or to perform an exclusive $\nu_{e}$ analysis. Given the recent advances in machine learning, we also plan to provide tools for turning our events into images that can be used in machine learning research.

\section{Data Preservation and Snowmass}

We intend to make all aspects of our data preservation product publicly available and documented sufficiently that a trained experimental neutrino physicist could, in principle, use them. However, we acknowledge that such a goal has not yet been realized by any modern neutrino experiments, and we worry that future analysis of our data may not be viable without some involvement from \minerva collaborators. As a practical matter, \minerva has always had the concept of limited authorship wherein temporary collaborators use our data to perform a specific measurement. We expect support for this analysis approach to extend beyond the current phase of our collaboration, so we expect our data to be useful to the community in the few-year time frame. For our data to be useful farther into the future, or for it to be usable without involvement from current \minerva collaborators, we believe that additional resources will be required. 

Because \minerva is a scintillator-based experiment, the disk footprint for storing the entirety of our data is small compared to some more recent neutrino interaction experiments. We currently expect the total size of our data set (FHC+RHC, LE+ME, Data+Simulation) will be $\sim10$ TB. The corresponding computational resources required to loop through these data vary with the complexity of the analysis, and in particular its dimensionality. For reference, a one-dimensional analysis filling $\mathcal{O}(10)$ histograms can run over the entire FHC ME data set in $\sim1$ hour using the default 2GB allocations of Fermilab's batch computing system. We believe that we will be able to maintain access to the data via Fermilab computing resources on the time scale of DUNE, but there is not an obvious longer-term storage option. Additionally, hosting the data at Fermilab will, presumably, restrict access to Fermilab users. We encourage the community to consider longer-term solutions for hosting not only preserved \minerva data but future data collected and preserved by other neutrino experiments.

\bibliographystyle{apsrev4-1}
\bibliography{refs} 

\begin{thebibliography}{55}%
\makeatletter
\providecommand \@ifxundefined [1]{%
 \@ifx{#1\undefined}
}%
\providecommand \@ifnum [1]{%
 \ifnum #1\expandafter \@firstoftwo
 \else \expandafter \@secondoftwo
 \fi
}%
\providecommand \@ifx [1]{%
 \ifx #1\expandafter \@firstoftwo
 \else \expandafter \@secondoftwo
 \fi
}%
\providecommand \natexlab [1]{#1}%
\providecommand \enquote  [1]{``#1''}%
\providecommand \bibnamefont  [1]{#1}%
\providecommand \bibfnamefont [1]{#1}%
\providecommand \citenamefont [1]{#1}%
\providecommand \href@noop [0]{\@secondoftwo}%
\providecommand \href [0]{\begingroup \@sanitize@url \@href}%
\providecommand \@href[1]{\@@startlink{#1}\@@href}%
\providecommand \@@href[1]{\endgroup#1\@@endlink}%
\providecommand \@sanitize@url [0]{\catcode `\\12\catcode `\$12\catcode
  `\&12\catcode `\#12\catcode `\^12\catcode `\_12\catcode `\%12\relax}%
\providecommand \@@startlink[1]{}%
\providecommand \@@endlink[0]{}%
\providecommand \url  [0]{\begingroup\@sanitize@url \@url }%
\providecommand \@url [1]{\endgroup\@href {#1}{\urlprefix }}%
\providecommand \urlprefix  [0]{URL }%
\providecommand \Eprint [0]{\href }%
\providecommand \doibase [0]{http://dx.doi.org/}%
\providecommand \selectlanguage [0]{\@gobble}%
\providecommand \bibinfo  [0]{\@secondoftwo}%
\providecommand \bibfield  [0]{\@secondoftwo}%
\providecommand \translation [1]{[#1]}%
\providecommand \BibitemOpen [0]{}%
\providecommand \bibitemStop [0]{}%
\providecommand \bibitemNoStop [0]{.\EOS\space}%
\providecommand \EOS [0]{\spacefactor3000\relax}%
\providecommand \BibitemShut  [1]{\csname bibitem#1\endcsname}%
\let\auto@bib@innerbib\@empty
%</preamble>
\bibitem [{\citenamefont {Barrand}\ \emph {et~al.}(2001)\citenamefont {Barrand}
  \emph {et~al.}}]{Barrand:2001ny}%
  \BibitemOpen
  \bibfield  {author} {\bibinfo {author} {\bibfnamefont {G.}~\bibnamefont
  {Barrand}} \emph {et~al.},\ }\href {\doibase 10.1016/S0010-4655(01)00254-5}
  {\bibfield  {journal} {\bibinfo  {journal} {Comput. Phys. Commun.}\ }\textbf
  {\bibinfo {volume} {140}},\ \bibinfo {pages} {45} (\bibinfo {year}
  {2001})}\BibitemShut {NoStop}%
\bibitem [{\citenamefont {Acciarri}\ \emph {et~al.}(2016)\citenamefont
  {Acciarri} \emph {et~al.}}]{Acciarri:2016crz}%
  \BibitemOpen
  \bibfield  {author} {\bibinfo {author} {\bibfnamefont {R.}~\bibnamefont
  {Acciarri}} \emph {et~al.} (\bibinfo {collaboration} {DUNE}),\ }\href@noop {}
  {\  (\bibinfo {year} {2016})},\ \Eprint {http://arxiv.org/abs/1601.05471}
  {arXiv:1601.05471 [physics.ins-det]} \BibitemShut {NoStop}%
\bibitem [{\citenamefont {Marshall}\ \emph {et~al.}(2020)\citenamefont
  {Marshall}, \citenamefont {McFarland},\ and\ \citenamefont
  {Wilkinson}}]{Marshall:2019vdy}%
  \BibitemOpen
  \bibfield  {author} {\bibinfo {author} {\bibfnamefont {C.~M.}\ \bibnamefont
  {Marshall}}, \bibinfo {author} {\bibfnamefont {K.~S.}\ \bibnamefont
  {McFarland}}, \ and\ \bibinfo {author} {\bibfnamefont {C.}~\bibnamefont
  {Wilkinson}},\ }\href {\doibase 10.1103/PhysRevD.101.032002} {\bibfield
  {journal} {\bibinfo  {journal} {Phys. Rev. D}\ }\textbf {\bibinfo {volume}
  {101}},\ \bibinfo {pages} {032002} (\bibinfo {year} {2020})},\ \Eprint
  {http://arxiv.org/abs/1910.10996} {arXiv:1910.10996 [hep-ex]} \BibitemShut
  {NoStop}%
\bibitem [{\citenamefont {Filkins}\ \emph {et~al.}(2020)\citenamefont {Filkins}
  \emph {et~al.}}]{MINERvA:2020zzv}%
  \BibitemOpen
  \bibfield  {author} {\bibinfo {author} {\bibfnamefont {A.}~\bibnamefont
  {Filkins}} \emph {et~al.} (\bibinfo {collaboration} {MINERvA}),\ }\href
  {\doibase 10.1103/PhysRevD.101.112007} {\bibfield  {journal} {\bibinfo
  {journal} {Phys. Rev. D}\ }\textbf {\bibinfo {volume} {101}},\ \bibinfo
  {pages} {112007} (\bibinfo {year} {2020})},\ \Eprint
  {http://arxiv.org/abs/2002.12496} {arXiv:2002.12496 [hep-ex]} \BibitemShut
  {NoStop}%
\bibitem [{\citenamefont {Coplowe}\ \emph
  {et~al.}(2020{\natexlab{a}})\citenamefont {Coplowe} \emph
  {et~al.}}]{MINERvA:2020anu}%
  \BibitemOpen
  \bibfield  {author} {\bibinfo {author} {\bibfnamefont {D.}~\bibnamefont
  {Coplowe}} \emph {et~al.} (\bibinfo {collaboration} {MINERvA}),\ }\href
  {\doibase 10.1103/PhysRevD.102.072007} {\bibfield  {journal} {\bibinfo
  {journal} {Phys. Rev. D}\ }\textbf {\bibinfo {volume} {102}},\ \bibinfo
  {pages} {072007} (\bibinfo {year} {2020}{\natexlab{a}})},\ \Eprint
  {http://arxiv.org/abs/2002.05812} {arXiv:2002.05812 [hep-ex]} \BibitemShut
  {NoStop}%
\bibitem [{\citenamefont {Cai}\ \emph {et~al.}(2020{\natexlab{a}})\citenamefont
  {Cai} \emph {et~al.}}]{MINERvA:2019ope}%
  \BibitemOpen
  \bibfield  {author} {\bibinfo {author} {\bibfnamefont {T.}~\bibnamefont
  {Cai}} \emph {et~al.} (\bibinfo {collaboration} {MINERvA}),\ }\href {\doibase
  10.1103/PhysRevD.101.092001} {\bibfield  {journal} {\bibinfo  {journal}
  {Phys. Rev. D}\ }\textbf {\bibinfo {volume} {101}},\ \bibinfo {pages}
  {092001} (\bibinfo {year} {2020}{\natexlab{a}})},\ \Eprint
  {http://arxiv.org/abs/1910.08658} {arXiv:1910.08658 [hep-ex]} \BibitemShut
  {NoStop}%
\bibitem [{\citenamefont {Le}\ \emph {et~al.}(2019)\citenamefont {Le} \emph
  {et~al.}}]{MINERvA:2019rhx}%
  \BibitemOpen
  \bibfield  {author} {\bibinfo {author} {\bibfnamefont {T.}~\bibnamefont {Le}}
  \emph {et~al.} (\bibinfo {collaboration} {MINERvA}),\ }\href {\doibase
  10.1103/PhysRevD.100.052008} {\bibfield  {journal} {\bibinfo  {journal}
  {Phys. Rev. D}\ }\textbf {\bibinfo {volume} {100}},\ \bibinfo {pages}
  {052008} (\bibinfo {year} {2019})},\ \Eprint
  {http://arxiv.org/abs/1906.08300} {arXiv:1906.08300 [hep-ex]} \BibitemShut
  {NoStop}%
\bibitem [{\citenamefont {Elkins}\ \emph {et~al.}(2019)\citenamefont {Elkins}
  \emph {et~al.}}]{MINERvA:2019wqe}%
  \BibitemOpen
  \bibfield  {author} {\bibinfo {author} {\bibfnamefont {M.}~\bibnamefont
  {Elkins}} \emph {et~al.} (\bibinfo {collaboration} {MINERvA}),\ }\href
  {\doibase 10.1103/PhysRevD.100.052002} {\bibfield  {journal} {\bibinfo
  {journal} {Phys. Rev. D}\ }\textbf {\bibinfo {volume} {100}},\ \bibinfo
  {pages} {052002} (\bibinfo {year} {2019})},\ \Eprint
  {http://arxiv.org/abs/1901.04892} {arXiv:1901.04892 [hep-ex]} \BibitemShut
  {NoStop}%
\bibitem [{\citenamefont {Ruterbories}\ \emph {et~al.}(2019)\citenamefont
  {Ruterbories} \emph {et~al.}}]{MINERvA:2018hqn}%
  \BibitemOpen
  \bibfield  {author} {\bibinfo {author} {\bibfnamefont {D.}~\bibnamefont
  {Ruterbories}} \emph {et~al.} (\bibinfo {collaboration} {MINERvA}),\ }\href
  {\doibase 10.1103/PhysRevD.99.012004} {\bibfield  {journal} {\bibinfo
  {journal} {Phys. Rev. D}\ }\textbf {\bibinfo {volume} {99}},\ \bibinfo
  {pages} {012004} (\bibinfo {year} {2019})},\ \Eprint
  {http://arxiv.org/abs/1811.02774} {arXiv:1811.02774 [hep-ex]} \BibitemShut
  {NoStop}%
\bibitem [{\citenamefont {Lu}\ \emph {et~al.}(2018{\natexlab{a}})\citenamefont
  {Lu} \emph {et~al.}}]{MINERvA:2018hba}%
  \BibitemOpen
  \bibfield  {author} {\bibinfo {author} {\bibfnamefont {X.~G.}\ \bibnamefont
  {Lu}} \emph {et~al.} (\bibinfo {collaboration} {MINERvA}),\ }\href {\doibase
  10.1103/PhysRevLett.121.022504} {\bibfield  {journal} {\bibinfo  {journal}
  {Phys. Rev. Lett.}\ }\textbf {\bibinfo {volume} {121}},\ \bibinfo {pages}
  {022504} (\bibinfo {year} {2018}{\natexlab{a}})},\ \Eprint
  {http://arxiv.org/abs/1805.05486} {arXiv:1805.05486 [hep-ex]} \BibitemShut
  {NoStop}%
\bibitem [{\citenamefont {Gran}\ \emph {et~al.}(2018)\citenamefont {Gran} \emph
  {et~al.}}]{MINERvA:2018nab}%
  \BibitemOpen
  \bibfield  {author} {\bibinfo {author} {\bibfnamefont {R.}~\bibnamefont
  {Gran}} \emph {et~al.} (\bibinfo {collaboration} {MINERvA}),\ }\href
  {\doibase 10.1103/PhysRevLett.120.221805} {\bibfield  {journal} {\bibinfo
  {journal} {Phys. Rev. Lett.}\ }\textbf {\bibinfo {volume} {120}},\ \bibinfo
  {pages} {221805} (\bibinfo {year} {2018})},\ \Eprint
  {http://arxiv.org/abs/1803.09377} {arXiv:1803.09377 [hep-ex]} \BibitemShut
  {NoStop}%
\bibitem [{\citenamefont {Patrick}\ \emph {et~al.}(2018)\citenamefont {Patrick}
  \emph {et~al.}}]{MINERvA:2018vjb}%
  \BibitemOpen
  \bibfield  {author} {\bibinfo {author} {\bibfnamefont {C.~E.}\ \bibnamefont
  {Patrick}} \emph {et~al.} (\bibinfo {collaboration} {MINERvA}),\ }\href
  {\doibase 10.1103/PhysRevD.97.052002} {\bibfield  {journal} {\bibinfo
  {journal} {Phys. Rev. D}\ }\textbf {\bibinfo {volume} {97}},\ \bibinfo
  {pages} {052002} (\bibinfo {year} {2018})},\ \Eprint
  {http://arxiv.org/abs/1801.01197} {arXiv:1801.01197 [hep-ex]} \BibitemShut
  {NoStop}%
\bibitem [{\citenamefont {Mislivec}\ \emph {et~al.}(2018)\citenamefont
  {Mislivec} \emph {et~al.}}]{MINERvA:2017ipy}%
  \BibitemOpen
  \bibfield  {author} {\bibinfo {author} {\bibfnamefont {A.}~\bibnamefont
  {Mislivec}} \emph {et~al.} (\bibinfo {collaboration} {MINERvA}),\ }\href
  {\doibase 10.1103/PhysRevD.97.032014} {\bibfield  {journal} {\bibinfo
  {journal} {Phys. Rev. D}\ }\textbf {\bibinfo {volume} {97}},\ \bibinfo
  {pages} {032014} (\bibinfo {year} {2018})},\ \Eprint
  {http://arxiv.org/abs/1711.01178} {arXiv:1711.01178 [hep-ex]} \BibitemShut
  {NoStop}%
\bibitem [{\citenamefont {Altinok}\ \emph {et~al.}(2017)\citenamefont {Altinok}
  \emph {et~al.}}]{MINERvA:2017okh}%
  \BibitemOpen
  \bibfield  {author} {\bibinfo {author} {\bibfnamefont {O.}~\bibnamefont
  {Altinok}} \emph {et~al.} (\bibinfo {collaboration} {MINERvA}),\ }\href
  {\doibase 10.1103/PhysRevD.96.072003} {\bibfield  {journal} {\bibinfo
  {journal} {Phys. Rev. D}\ }\textbf {\bibinfo {volume} {96}},\ \bibinfo
  {pages} {072003} (\bibinfo {year} {2017})},\ \Eprint
  {http://arxiv.org/abs/1708.03723} {arXiv:1708.03723 [hep-ex]} \BibitemShut
  {NoStop}%
\bibitem [{\citenamefont {Betancourt}\ \emph {et~al.}(2017)\citenamefont
  {Betancourt} \emph {et~al.}}]{MINERvA:2017dzh}%
  \BibitemOpen
  \bibfield  {author} {\bibinfo {author} {\bibfnamefont {M.}~\bibnamefont
  {Betancourt}} \emph {et~al.} (\bibinfo {collaboration} {MINERvA}),\ }\href
  {\doibase 10.1103/PhysRevLett.119.082001} {\bibfield  {journal} {\bibinfo
  {journal} {Phys. Rev. Lett.}\ }\textbf {\bibinfo {volume} {119}},\ \bibinfo
  {pages} {082001} (\bibinfo {year} {2017})},\ \Eprint
  {http://arxiv.org/abs/1705.03791} {arXiv:1705.03791 [hep-ex]} \BibitemShut
  {NoStop}%
\bibitem [{\citenamefont {Ren}\ \emph {et~al.}(2017)\citenamefont {Ren} \emph
  {et~al.}}]{MINERvA:2017ozn}%
  \BibitemOpen
  \bibfield  {author} {\bibinfo {author} {\bibfnamefont {L.}~\bibnamefont
  {Ren}} \emph {et~al.} (\bibinfo {collaboration} {MINERvA}),\ }\href {\doibase
  10.1103/PhysRevD.95.072009} {\bibfield  {journal} {\bibinfo  {journal} {Phys.
  Rev. D}\ }\textbf {\bibinfo {volume} {95}},\ \bibinfo {pages} {072009}
  (\bibinfo {year} {2017})},\ \bibinfo {note} {[Addendum: Phys.Rev.D 97, 019902
  (2018)]},\ \Eprint {http://arxiv.org/abs/1701.04857} {arXiv:1701.04857
  [hep-ex]} \BibitemShut {NoStop}%
\bibitem [{\citenamefont {Marshall}\ \emph {et~al.}(2017)\citenamefont
  {Marshall} \emph {et~al.}}]{MINERvA:2016ymg}%
  \BibitemOpen
  \bibfield  {author} {\bibinfo {author} {\bibfnamefont {C.~M.}\ \bibnamefont
  {Marshall}} \emph {et~al.} (\bibinfo {collaboration} {MINERvA}),\ }\href
  {\doibase 10.1103/PhysRevLett.119.011802} {\bibfield  {journal} {\bibinfo
  {journal} {Phys. Rev. Lett.}\ }\textbf {\bibinfo {volume} {119}},\ \bibinfo
  {pages} {011802} (\bibinfo {year} {2017})},\ \Eprint
  {http://arxiv.org/abs/1611.02224} {arXiv:1611.02224 [hep-ex]} \BibitemShut
  {NoStop}%
\bibitem [{\citenamefont {Devan}\ \emph {et~al.}(2016)\citenamefont {Devan}
  \emph {et~al.}}]{MINERvA:2016ing}%
  \BibitemOpen
  \bibfield  {author} {\bibinfo {author} {\bibfnamefont {J.}~\bibnamefont
  {Devan}} \emph {et~al.} (\bibinfo {collaboration} {MINERvA}),\ }\href
  {\doibase 10.1103/PhysRevD.94.112007} {\bibfield  {journal} {\bibinfo
  {journal} {Phys. Rev. D}\ }\textbf {\bibinfo {volume} {94}},\ \bibinfo
  {pages} {112007} (\bibinfo {year} {2016})},\ \Eprint
  {http://arxiv.org/abs/1610.04746} {arXiv:1610.04746 [hep-ex]} \BibitemShut
  {NoStop}%
\bibitem [{\citenamefont {Wang}\ \emph {et~al.}(2016)\citenamefont {Wang} \emph
  {et~al.}}]{MINERvA:2016cun}%
  \BibitemOpen
  \bibfield  {author} {\bibinfo {author} {\bibfnamefont {Z.}~\bibnamefont
  {Wang}} \emph {et~al.} (\bibinfo {collaboration} {MINERvA}),\ }\href
  {\doibase 10.1103/PhysRevLett.117.061802} {\bibfield  {journal} {\bibinfo
  {journal} {Phys. Rev. Lett.}\ }\textbf {\bibinfo {volume} {117}},\ \bibinfo
  {pages} {061802} (\bibinfo {year} {2016})},\ \Eprint
  {http://arxiv.org/abs/1606.08890} {arXiv:1606.08890 [hep-ex]} \BibitemShut
  {NoStop}%
\bibitem [{\citenamefont {McGivern}\ \emph {et~al.}(2016)\citenamefont
  {McGivern} \emph {et~al.}}]{MINERvA:2016sfc}%
  \BibitemOpen
  \bibfield  {author} {\bibinfo {author} {\bibfnamefont {C.~L.}\ \bibnamefont
  {McGivern}} \emph {et~al.} (\bibinfo {collaboration} {MINERvA}),\ }\href
  {\doibase 10.1103/PhysRevD.94.052005} {\bibfield  {journal} {\bibinfo
  {journal} {Phys. Rev. D}\ }\textbf {\bibinfo {volume} {94}},\ \bibinfo
  {pages} {052005} (\bibinfo {year} {2016})},\ \Eprint
  {http://arxiv.org/abs/1606.07127} {arXiv:1606.07127 [hep-ex]} \BibitemShut
  {NoStop}%
\bibitem [{\citenamefont {Marshall}\ \emph {et~al.}(2016)\citenamefont
  {Marshall} \emph {et~al.}}]{MINERvA:2016zyp}%
  \BibitemOpen
  \bibfield  {author} {\bibinfo {author} {\bibfnamefont {C.~M.}\ \bibnamefont
  {Marshall}} \emph {et~al.} (\bibinfo {collaboration} {MINERvA}),\ }\href
  {\doibase 10.1103/PhysRevD.94.012002} {\bibfield  {journal} {\bibinfo
  {journal} {Phys. Rev. D}\ }\textbf {\bibinfo {volume} {94}},\ \bibinfo
  {pages} {012002} (\bibinfo {year} {2016})},\ \Eprint
  {http://arxiv.org/abs/1604.03920} {arXiv:1604.03920 [hep-ex]} \BibitemShut
  {NoStop}%
\bibitem [{\citenamefont {Wolcott}\ \emph
  {et~al.}(2016{\natexlab{a}})\citenamefont {Wolcott} \emph
  {et~al.}}]{MINERvA:2016uck}%
  \BibitemOpen
  \bibfield  {author} {\bibinfo {author} {\bibfnamefont {J.}~\bibnamefont
  {Wolcott}} \emph {et~al.} (\bibinfo {collaboration} {MINERvA}),\ }\href
  {\doibase 10.1103/PhysRevLett.117.111801} {\bibfield  {journal} {\bibinfo
  {journal} {Phys. Rev. Lett.}\ }\textbf {\bibinfo {volume} {117}},\ \bibinfo
  {pages} {111801} (\bibinfo {year} {2016}{\natexlab{a}})},\ \Eprint
  {http://arxiv.org/abs/1604.01728} {arXiv:1604.01728 [hep-ex]} \BibitemShut
  {NoStop}%
\bibitem [{\citenamefont {Mousseau}\ \emph {et~al.}(2016)\citenamefont
  {Mousseau} \emph {et~al.}}]{MINERvA:2016oql}%
  \BibitemOpen
  \bibfield  {author} {\bibinfo {author} {\bibfnamefont {J.}~\bibnamefont
  {Mousseau}} \emph {et~al.} (\bibinfo {collaboration} {MINERvA}),\ }\href
  {\doibase 10.1103/PhysRevD.93.071101} {\bibfield  {journal} {\bibinfo
  {journal} {Phys. Rev. D}\ }\textbf {\bibinfo {volume} {93}},\ \bibinfo
  {pages} {071101} (\bibinfo {year} {2016})},\ \Eprint
  {http://arxiv.org/abs/1601.06313} {arXiv:1601.06313 [hep-ex]} \BibitemShut
  {NoStop}%
\bibitem [{\citenamefont {Park}\ \emph {et~al.}(2016)\citenamefont {Park} \emph
  {et~al.}}]{MINERvA:2015nqi}%
  \BibitemOpen
  \bibfield  {author} {\bibinfo {author} {\bibfnamefont {J.}~\bibnamefont
  {Park}} \emph {et~al.} (\bibinfo {collaboration} {MINERvA}),\ }\href
  {\doibase 10.1103/PhysRevD.93.112007} {\bibfield  {journal} {\bibinfo
  {journal} {Phys. Rev. D}\ }\textbf {\bibinfo {volume} {93}},\ \bibinfo
  {pages} {112007} (\bibinfo {year} {2016})},\ \Eprint
  {http://arxiv.org/abs/1512.07699} {arXiv:1512.07699 [physics.ins-det]}
  \BibitemShut {NoStop}%
\bibitem [{\citenamefont {Rodrigues}\ \emph {et~al.}(2016)\citenamefont
  {Rodrigues} \emph {et~al.}}]{MINERvA:2015ydy}%
  \BibitemOpen
  \bibfield  {author} {\bibinfo {author} {\bibfnamefont {P.~A.}\ \bibnamefont
  {Rodrigues}} \emph {et~al.} (\bibinfo {collaboration} {MINERvA}),\ }\href
  {\doibase 10.1103/PhysRevLett.116.071802} {\bibfield  {journal} {\bibinfo
  {journal} {Phys. Rev. Lett.}\ }\textbf {\bibinfo {volume} {116}},\ \bibinfo
  {pages} {071802} (\bibinfo {year} {2016})},\ \bibinfo {note} {[Addendum:
  Phys.Rev.Lett. 121, 209902 (2018)]},\ \Eprint
  {http://arxiv.org/abs/1511.05944} {arXiv:1511.05944 [hep-ex]} \BibitemShut
  {NoStop}%
\bibitem [{\citenamefont {Wolcott}\ \emph
  {et~al.}(2016{\natexlab{b}})\citenamefont {Wolcott} \emph
  {et~al.}}]{MINERvA:2015jih}%
  \BibitemOpen
  \bibfield  {author} {\bibinfo {author} {\bibfnamefont {J.}~\bibnamefont
  {Wolcott}} \emph {et~al.} (\bibinfo {collaboration} {MINERvA}),\ }\href
  {\doibase 10.1103/PhysRevLett.116.081802} {\bibfield  {journal} {\bibinfo
  {journal} {Phys. Rev. Lett.}\ }\textbf {\bibinfo {volume} {116}},\ \bibinfo
  {pages} {081802} (\bibinfo {year} {2016}{\natexlab{b}})},\ \Eprint
  {http://arxiv.org/abs/1509.05729} {arXiv:1509.05729 [hep-ex]} \BibitemShut
  {NoStop}%
\bibitem [{\citenamefont {Le}\ \emph {et~al.}(2015)\citenamefont {Le} \emph
  {et~al.}}]{MINERvA:2015slz}%
  \BibitemOpen
  \bibfield  {author} {\bibinfo {author} {\bibfnamefont {T.}~\bibnamefont {Le}}
  \emph {et~al.} (\bibinfo {collaboration} {MINERvA}),\ }\href {\doibase
  10.1016/j.physletb.2015.07.039} {\bibfield  {journal} {\bibinfo  {journal}
  {Phys. Lett. B}\ }\textbf {\bibinfo {volume} {749}},\ \bibinfo {pages} {130}
  (\bibinfo {year} {2015})},\ \Eprint {http://arxiv.org/abs/1503.02107}
  {arXiv:1503.02107 [hep-ex]} \BibitemShut {NoStop}%
\bibitem [{\citenamefont {Walton}\ \emph {et~al.}(2015)\citenamefont {Walton}
  \emph {et~al.}}]{MINERvA:2014ypj}%
  \BibitemOpen
  \bibfield  {author} {\bibinfo {author} {\bibfnamefont {T.}~\bibnamefont
  {Walton}} \emph {et~al.} (\bibinfo {collaboration} {MINERvA}),\ }\href
  {\doibase 10.1103/PhysRevD.91.071301} {\bibfield  {journal} {\bibinfo
  {journal} {Phys. Rev. D}\ }\textbf {\bibinfo {volume} {91}},\ \bibinfo
  {pages} {071301} (\bibinfo {year} {2015})},\ \Eprint
  {http://arxiv.org/abs/1409.4497} {arXiv:1409.4497 [hep-ex]} \BibitemShut
  {NoStop}%
\bibitem [{\citenamefont {Higuera}\ \emph {et~al.}(2014)\citenamefont {Higuera}
  \emph {et~al.}}]{MINERvA:2014ani}%
  \BibitemOpen
  \bibfield  {author} {\bibinfo {author} {\bibfnamefont {A.}~\bibnamefont
  {Higuera}} \emph {et~al.} (\bibinfo {collaboration} {MINERvA}),\ }\href
  {\doibase 10.1103/PhysRevLett.113.261802} {\bibfield  {journal} {\bibinfo
  {journal} {Phys. Rev. Lett.}\ }\textbf {\bibinfo {volume} {113}},\ \bibinfo
  {pages} {261802} (\bibinfo {year} {2014})},\ \Eprint
  {http://arxiv.org/abs/1409.3835} {arXiv:1409.3835 [hep-ex]} \BibitemShut
  {NoStop}%
\bibitem [{\citenamefont {Eberly}\ \emph {et~al.}(2015)\citenamefont {Eberly}
  \emph {et~al.}}]{MINERvA:2014ogb}%
  \BibitemOpen
  \bibfield  {author} {\bibinfo {author} {\bibfnamefont {B.}~\bibnamefont
  {Eberly}} \emph {et~al.} (\bibinfo {collaboration} {MINERvA}),\ }\href
  {\doibase 10.1103/PhysRevD.92.092008} {\bibfield  {journal} {\bibinfo
  {journal} {Phys. Rev. D}\ }\textbf {\bibinfo {volume} {92}},\ \bibinfo
  {pages} {092008} (\bibinfo {year} {2015})},\ \Eprint
  {http://arxiv.org/abs/1406.6415} {arXiv:1406.6415 [hep-ex]} \BibitemShut
  {NoStop}%
\bibitem [{\citenamefont {Tice}\ \emph {et~al.}(2014)\citenamefont {Tice} \emph
  {et~al.}}]{MINERvA:2014rdw}%
  \BibitemOpen
  \bibfield  {author} {\bibinfo {author} {\bibfnamefont {B.~G.}\ \bibnamefont
  {Tice}} \emph {et~al.} (\bibinfo {collaboration} {MINERvA}),\ }\href
  {\doibase 10.1103/PhysRevLett.112.231801} {\bibfield  {journal} {\bibinfo
  {journal} {Phys. Rev. Lett.}\ }\textbf {\bibinfo {volume} {112}},\ \bibinfo
  {pages} {231801} (\bibinfo {year} {2014})},\ \Eprint
  {http://arxiv.org/abs/1403.2103} {arXiv:1403.2103 [hep-ex]} \BibitemShut
  {NoStop}%
\bibitem [{\citenamefont {Fields}\ \emph {et~al.}(2013)\citenamefont {Fields}
  \emph {et~al.}}]{MINERvA:2013bcy}%
  \BibitemOpen
  \bibfield  {author} {\bibinfo {author} {\bibfnamefont {L.}~\bibnamefont
  {Fields}} \emph {et~al.} (\bibinfo {collaboration} {MINERvA}),\ }\href
  {\doibase 10.1103/PhysRevLett.111.022501} {\bibfield  {journal} {\bibinfo
  {journal} {Phys. Rev. Lett.}\ }\textbf {\bibinfo {volume} {111}},\ \bibinfo
  {pages} {022501} (\bibinfo {year} {2013})},\ \Eprint
  {http://arxiv.org/abs/1305.2234} {arXiv:1305.2234 [hep-ex]} \BibitemShut
  {NoStop}%
\bibitem [{\citenamefont {Fiorentini}\ \emph {et~al.}(2013)\citenamefont
  {Fiorentini} \emph {et~al.}}]{MINERvA:2013kdn}%
  \BibitemOpen
  \bibfield  {author} {\bibinfo {author} {\bibfnamefont {G.~A.}\ \bibnamefont
  {Fiorentini}} \emph {et~al.} (\bibinfo {collaboration} {MINERvA}),\ }\href
  {\doibase 10.1103/PhysRevLett.111.022502} {\bibfield  {journal} {\bibinfo
  {journal} {Phys. Rev. Lett.}\ }\textbf {\bibinfo {volume} {111}},\ \bibinfo
  {pages} {022502} (\bibinfo {year} {2013})},\ \Eprint
  {http://arxiv.org/abs/1305.2243} {arXiv:1305.2243 [hep-ex]} \BibitemShut
  {NoStop}%
\bibitem [{\citenamefont {Ruterbories}\ \emph {et~al.}(2022)\citenamefont
  {Ruterbories} \emph {et~al.}}]{MINERvA:2022mnw}%
  \BibitemOpen
  \bibfield  {author} {\bibinfo {author} {\bibfnamefont {D.}~\bibnamefont
  {Ruterbories}} \emph {et~al.} (\bibinfo {collaboration} {MINERvA}),\ }\href
  {\doibase 10.1103/PhysRevLett.129.021803} {\bibfield  {journal} {\bibinfo
  {journal} {Phys. Rev. Lett.}\ }\textbf {\bibinfo {volume} {129}},\ \bibinfo
  {pages} {021803} (\bibinfo {year} {2022})},\ \Eprint
  {http://arxiv.org/abs/2203.08022} {arXiv:2203.08022 [hep-ex]} \BibitemShut
  {NoStop}%
\bibitem [{\citenamefont {Ascencio}\ \emph {et~al.}(2022)\citenamefont
  {Ascencio} \emph {et~al.}}]{MINERvA:2021wjs}%
  \BibitemOpen
  \bibfield  {author} {\bibinfo {author} {\bibfnamefont {M.~V.}\ \bibnamefont
  {Ascencio}} \emph {et~al.} (\bibinfo {collaboration} {MINERvA}),\ }\href
  {\doibase 10.1103/PhysRevD.106.032001} {\bibfield  {journal} {\bibinfo
  {journal} {Phys. Rev. D}\ }\textbf {\bibinfo {volume} {106}},\ \bibinfo
  {pages} {032001} (\bibinfo {year} {2022})},\ \Eprint
  {http://arxiv.org/abs/2110.13372} {arXiv:2110.13372 [hep-ex]} \BibitemShut
  {NoStop}%
\bibitem [{\citenamefont {Ruterbories}\ \emph
  {et~al.}(2021{\natexlab{a}})\citenamefont {Ruterbories} \emph
  {et~al.}}]{MINERvA:2021dhf}%
  \BibitemOpen
  \bibfield  {author} {\bibinfo {author} {\bibfnamefont {D.}~\bibnamefont
  {Ruterbories}} \emph {et~al.} (\bibinfo {collaboration} {MINERvA}),\ }\href
  {\doibase 10.1103/PhysRevD.104.092010} {\bibfield  {journal} {\bibinfo
  {journal} {Phys. Rev. D}\ }\textbf {\bibinfo {volume} {104}},\ \bibinfo
  {pages} {092010} (\bibinfo {year} {2021}{\natexlab{a}})},\ \Eprint
  {http://arxiv.org/abs/2107.01059} {arXiv:2107.01059 [hep-ex]} \BibitemShut
  {NoStop}%
\bibitem [{\citenamefont {Ruterbories}\ \emph
  {et~al.}(2021{\natexlab{b}})\citenamefont {Ruterbories} \emph
  {et~al.}}]{MINERvA:2021owq}%
  \BibitemOpen
  \bibfield  {author} {\bibinfo {author} {\bibfnamefont {D.}~\bibnamefont
  {Ruterbories}} \emph {et~al.} (\bibinfo {collaboration} {MINERvA}),\ }\href
  {\doibase 10.1103/PhysRevD.104.092007} {\bibfield  {journal} {\bibinfo
  {journal} {Phys. Rev. D}\ }\textbf {\bibinfo {volume} {104}},\ \bibinfo
  {pages} {092007} (\bibinfo {year} {2021}{\natexlab{b}})},\ \Eprint
  {http://arxiv.org/abs/2106.16210} {arXiv:2106.16210 [hep-ex]} \BibitemShut
  {NoStop}%
\bibitem [{\citenamefont {Carneiro}\ \emph {et~al.}(2020)\citenamefont
  {Carneiro} \emph {et~al.}}]{MINERvA:2019gsf}%
  \BibitemOpen
  \bibfield  {author} {\bibinfo {author} {\bibfnamefont {M.~F.}\ \bibnamefont
  {Carneiro}} \emph {et~al.} (\bibinfo {collaboration} {MINERvA}),\ }\href
  {\doibase 10.1103/PhysRevLett.124.121801} {\bibfield  {journal} {\bibinfo
  {journal} {Phys. Rev. Lett.}\ }\textbf {\bibinfo {volume} {124}},\ \bibinfo
  {pages} {121801} (\bibinfo {year} {2020})},\ \Eprint
  {http://arxiv.org/abs/1912.09890} {arXiv:1912.09890 [hep-ex]} \BibitemShut
  {NoStop}%
\bibitem [{\citenamefont {Valencia}\ \emph {et~al.}(2019)\citenamefont
  {Valencia} \emph {et~al.}}]{MINERvA:2019hhc}%
  \BibitemOpen
  \bibfield  {author} {\bibinfo {author} {\bibfnamefont {E.}~\bibnamefont
  {Valencia}} \emph {et~al.} (\bibinfo {collaboration} {MINERvA}),\ }\href
  {\doibase 10.1103/PhysRevD.100.092001} {\bibfield  {journal} {\bibinfo
  {journal} {Phys. Rev. D}\ }\textbf {\bibinfo {volume} {100}},\ \bibinfo
  {pages} {092001} (\bibinfo {year} {2019})},\ \Eprint
  {http://arxiv.org/abs/1906.00111} {arXiv:1906.00111 [hep-ex]} \BibitemShut
  {NoStop}%
\bibitem [{\citenamefont {Lu}\ \emph {et~al.}(2016)\citenamefont {Lu},
  \citenamefont {Pickering}, \citenamefont {Dolan}, \citenamefont {Barr},
  \citenamefont {Coplowe}, \citenamefont {Uchida}, \citenamefont {Wark},
  \citenamefont {Wascko}, \citenamefont {Weber},\ and\ \citenamefont
  {Yuan}}]{Lu:2015tcr}%
  \BibitemOpen
  \bibfield  {author} {\bibinfo {author} {\bibfnamefont {X.~G.}\ \bibnamefont
  {Lu}}, \bibinfo {author} {\bibfnamefont {L.}~\bibnamefont {Pickering}},
  \bibinfo {author} {\bibfnamefont {S.}~\bibnamefont {Dolan}}, \bibinfo
  {author} {\bibfnamefont {G.}~\bibnamefont {Barr}}, \bibinfo {author}
  {\bibfnamefont {D.}~\bibnamefont {Coplowe}}, \bibinfo {author} {\bibfnamefont
  {Y.}~\bibnamefont {Uchida}}, \bibinfo {author} {\bibfnamefont
  {D.}~\bibnamefont {Wark}}, \bibinfo {author} {\bibfnamefont {M.}~\bibnamefont
  {Wascko}}, \bibinfo {author} {\bibfnamefont {A.}~\bibnamefont {Weber}}, \
  and\ \bibinfo {author} {\bibfnamefont {T.}~\bibnamefont {Yuan}},\ }\href
  {\doibase 10.1103/PhysRevC.94.015503} {\bibfield  {journal} {\bibinfo
  {journal} {Phys. Rev. C}\ }\textbf {\bibinfo {volume} {94}},\ \bibinfo
  {pages} {015503} (\bibinfo {year} {2016})},\ \Eprint
  {http://arxiv.org/abs/1512.05748} {arXiv:1512.05748 [nucl-th]} \BibitemShut
  {NoStop}%
\bibitem [{\citenamefont {Furmanski}\ and\ \citenamefont
  {Sobczyk}(2017)}]{Furmanski:2016wqo}%
  \BibitemOpen
  \bibfield  {author} {\bibinfo {author} {\bibfnamefont {A.~P.}\ \bibnamefont
  {Furmanski}}\ and\ \bibinfo {author} {\bibfnamefont {J.~T.}\ \bibnamefont
  {Sobczyk}},\ }\href {\doibase 10.1103/PhysRevC.95.065501} {\bibfield
  {journal} {\bibinfo  {journal} {Phys. Rev. C}\ }\textbf {\bibinfo {volume}
  {95}},\ \bibinfo {pages} {065501} (\bibinfo {year} {2017})},\ \Eprint
  {http://arxiv.org/abs/1609.03530} {arXiv:1609.03530 [hep-ex]} \BibitemShut
  {NoStop}%
\bibitem [{\citenamefont {Abe}\ \emph {et~al.}(2018)\citenamefont {Abe} \emph
  {et~al.}}]{Abe:2018pwo}%
  \BibitemOpen
  \bibfield  {author} {\bibinfo {author} {\bibfnamefont {K.}~\bibnamefont
  {Abe}} \emph {et~al.} (\bibinfo {collaboration} {T2K}),\ }\href {\doibase
  10.1103/PhysRevD.98.032003} {\bibfield  {journal} {\bibinfo  {journal} {Phys.
  Rev. D}\ }\textbf {\bibinfo {volume} {98}},\ \bibinfo {pages} {032003}
  (\bibinfo {year} {2018})},\ \Eprint {http://arxiv.org/abs/1802.05078}
  {arXiv:1802.05078 [hep-ex]} \BibitemShut {NoStop}%
\bibitem [{\citenamefont {Dolan}\ \emph {et~al.}(2018)\citenamefont {Dolan},
  \citenamefont {Mosel}, \citenamefont {Gallmeister}, \citenamefont
  {Pickering},\ and\ \citenamefont {Bolognesi}}]{Dolan:2018sbb}%
  \BibitemOpen
  \bibfield  {author} {\bibinfo {author} {\bibfnamefont {S.}~\bibnamefont
  {Dolan}}, \bibinfo {author} {\bibfnamefont {U.}~\bibnamefont {Mosel}},
  \bibinfo {author} {\bibfnamefont {K.}~\bibnamefont {Gallmeister}}, \bibinfo
  {author} {\bibfnamefont {L.}~\bibnamefont {Pickering}}, \ and\ \bibinfo
  {author} {\bibfnamefont {S.}~\bibnamefont {Bolognesi}},\ }\href {\doibase
  10.1103/PhysRevC.98.045502} {\bibfield  {journal} {\bibinfo  {journal} {Phys.
  Rev. C}\ }\textbf {\bibinfo {volume} {98}},\ \bibinfo {pages} {045502}
  (\bibinfo {year} {2018})},\ \Eprint {http://arxiv.org/abs/1804.09488}
  {arXiv:1804.09488 [hep-ex]} \BibitemShut {NoStop}%
\bibitem [{\citenamefont {Lu}\ \emph {et~al.}(2018{\natexlab{b}})\citenamefont
  {Lu} \emph {et~al.}}]{Lu:2018stk}%
  \BibitemOpen
  \bibfield  {author} {\bibinfo {author} {\bibfnamefont {X.}~\bibnamefont {Lu}}
  \emph {et~al.} (\bibinfo {collaboration} {MINERvA}),\ }\href {\doibase
  10.1103/PhysRevLett.121.022504} {\bibfield  {journal} {\bibinfo  {journal}
  {Phys. Rev. Lett.}\ }\textbf {\bibinfo {volume} {121}},\ \bibinfo {pages}
  {022504} (\bibinfo {year} {2018}{\natexlab{b}})},\ \Eprint
  {http://arxiv.org/abs/1805.05486} {arXiv:1805.05486 [hep-ex]} \BibitemShut
  {NoStop}%
\bibitem [{\citenamefont {Dolan}(2018)}]{Dolan:2018zye}%
  \BibitemOpen
  \bibfield  {author} {\bibinfo {author} {\bibfnamefont {S.}~\bibnamefont
  {Dolan}},\ }\href@noop {} {\  (\bibinfo {year} {2018})},\ \Eprint
  {http://arxiv.org/abs/1810.06043} {arXiv:1810.06043 [hep-ex]} \BibitemShut
  {NoStop}%
\bibitem [{\citenamefont {Lu}\ and\ \citenamefont
  {Sobczyk}(2019)}]{Lu:2019nmf}%
  \BibitemOpen
  \bibfield  {author} {\bibinfo {author} {\bibfnamefont {X.}~\bibnamefont
  {Lu}}\ and\ \bibinfo {author} {\bibfnamefont {J.~T.}\ \bibnamefont
  {Sobczyk}},\ }\href {\doibase 10.1103/PhysRevC.99.055504} {\bibfield
  {journal} {\bibinfo  {journal} {Phys. Rev. C}\ }\textbf {\bibinfo {volume}
  {99}},\ \bibinfo {pages} {055504} (\bibinfo {year} {2019})},\ \Eprint
  {http://arxiv.org/abs/1901.06411} {arXiv:1901.06411 [hep-ph]} \BibitemShut
  {NoStop}%
\bibitem [{\citenamefont {Harewood}\ and\ \citenamefont
  {Gran}(2019)}]{Harewood:2019rzy}%
  \BibitemOpen
  \bibfield  {author} {\bibinfo {author} {\bibfnamefont {L.}~\bibnamefont
  {Harewood}}\ and\ \bibinfo {author} {\bibfnamefont {R.}~\bibnamefont
  {Gran}},\ }\href@noop {} {\  (\bibinfo {year} {2019})},\ \Eprint
  {http://arxiv.org/abs/1906.10576} {arXiv:1906.10576 [hep-ex]} \BibitemShut
  {NoStop}%
\bibitem [{\citenamefont {Cai}\ \emph {et~al.}(2019)\citenamefont {Cai},
  \citenamefont {Lu},\ and\ \citenamefont {Ruterbories}}]{Cai:2019jzk}%
  \BibitemOpen
  \bibfield  {author} {\bibinfo {author} {\bibfnamefont {T.}~\bibnamefont
  {Cai}}, \bibinfo {author} {\bibfnamefont {X.}~\bibnamefont {Lu}}, \ and\
  \bibinfo {author} {\bibfnamefont {D.}~\bibnamefont {Ruterbories}},\ }\href
  {\doibase 10.1103/PhysRevD.100.073010} {\bibfield  {journal} {\bibinfo
  {journal} {Phys. Rev. D}\ }\textbf {\bibinfo {volume} {100}},\ \bibinfo
  {pages} {073010} (\bibinfo {year} {2019})},\ \Eprint
  {http://arxiv.org/abs/1907.11212} {arXiv:1907.11212 [hep-ex]} \BibitemShut
  {NoStop}%
\bibitem [{\citenamefont {Cai}\ \emph {et~al.}(2020{\natexlab{b}})\citenamefont
  {Cai} \emph {et~al.}}]{Cai:2019hpx}%
  \BibitemOpen
  \bibfield  {author} {\bibinfo {author} {\bibfnamefont {T.}~\bibnamefont
  {Cai}} \emph {et~al.} (\bibinfo {collaboration} {MINERvA}),\ }\href {\doibase
  10.1103/PhysRevD.101.092001} {\bibfield  {journal} {\bibinfo  {journal}
  {Phys. Rev. D}\ }\textbf {\bibinfo {volume} {101}},\ \bibinfo {pages}
  {092001} (\bibinfo {year} {2020}{\natexlab{b}})},\ \Eprint
  {http://arxiv.org/abs/1910.08658} {arXiv:1910.08658 [hep-ex]} \BibitemShut
  {NoStop}%
\bibitem [{\citenamefont {Coplowe}\ \emph
  {et~al.}(2020{\natexlab{b}})\citenamefont {Coplowe} \emph
  {et~al.}}]{Coplowe:2020yea}%
  \BibitemOpen
  \bibfield  {author} {\bibinfo {author} {\bibfnamefont {D.}~\bibnamefont
  {Coplowe}} \emph {et~al.} (\bibinfo {collaboration} {MINERvA}),\ }\href@noop
  {} {\  (\bibinfo {year} {2020}{\natexlab{b}})},\ \Eprint
  {http://arxiv.org/abs/2002.05812} {arXiv:2002.05812 [hep-ex]} \BibitemShut
  {NoStop}%
\bibitem [{\citenamefont {Lu}\ \emph {et~al.}(2015)\citenamefont {Lu},
  \citenamefont {Coplowe}, \citenamefont {Shah}, \citenamefont {Barr},
  \citenamefont {Wark},\ and\ \citenamefont {Weber}}]{Lu:2015hea}%
  \BibitemOpen
  \bibfield  {author} {\bibinfo {author} {\bibfnamefont {X.-G.}\ \bibnamefont
  {Lu}}, \bibinfo {author} {\bibfnamefont {D.}~\bibnamefont {Coplowe}},
  \bibinfo {author} {\bibfnamefont {R.}~\bibnamefont {Shah}}, \bibinfo {author}
  {\bibfnamefont {G.}~\bibnamefont {Barr}}, \bibinfo {author} {\bibfnamefont
  {D.}~\bibnamefont {Wark}}, \ and\ \bibinfo {author} {\bibfnamefont
  {A.}~\bibnamefont {Weber}},\ }\href {\doibase 10.1103/PhysRevD.92.051302}
  {\bibfield  {journal} {\bibinfo  {journal} {Phys. Rev. D}\ }\textbf {\bibinfo
  {volume} {92}},\ \bibinfo {pages} {051302} (\bibinfo {year} {2015})},\
  \Eprint {http://arxiv.org/abs/1507.00967} {arXiv:1507.00967 [hep-ex]}
  \BibitemShut {NoStop}%
\bibitem [{\citenamefont {Duyang}\ \emph {et~al.}(2018)\citenamefont {Duyang},
  \citenamefont {Guo}, \citenamefont {Mishra},\ and\ \citenamefont
  {Petti}}]{Duyang:2018lpe}%
  \BibitemOpen
  \bibfield  {author} {\bibinfo {author} {\bibfnamefont {H.}~\bibnamefont
  {Duyang}}, \bibinfo {author} {\bibfnamefont {B.}~\bibnamefont {Guo}},
  \bibinfo {author} {\bibfnamefont {S.}~\bibnamefont {Mishra}}, \ and\ \bibinfo
  {author} {\bibfnamefont {R.}~\bibnamefont {Petti}},\ }\href@noop {} {\
  (\bibinfo {year} {2018})},\ \Eprint {http://arxiv.org/abs/1809.08752}
  {arXiv:1809.08752 [hep-ph]} \BibitemShut {NoStop}%
\bibitem [{\citenamefont {Duyang}\ \emph {et~al.}(2019)\citenamefont {Duyang},
  \citenamefont {Guo}, \citenamefont {Mishra},\ and\ \citenamefont
  {Petti}}]{Duyang:2019prb}%
  \BibitemOpen
  \bibfield  {author} {\bibinfo {author} {\bibfnamefont {H.}~\bibnamefont
  {Duyang}}, \bibinfo {author} {\bibfnamefont {B.}~\bibnamefont {Guo}},
  \bibinfo {author} {\bibfnamefont {S.}~\bibnamefont {Mishra}}, \ and\ \bibinfo
  {author} {\bibfnamefont {R.}~\bibnamefont {Petti}},\ }\href {\doibase
  10.1016/j.physletb.2019.06.003} {\bibfield  {journal} {\bibinfo  {journal}
  {Phys. Lett. B}\ }\textbf {\bibinfo {volume} {795}},\ \bibinfo {pages} {424}
  (\bibinfo {year} {2019})},\ \Eprint {http://arxiv.org/abs/1902.09480}
  {arXiv:1902.09480 [hep-ph]} \BibitemShut {NoStop}%
\bibitem [{\citenamefont {Munteanu}\ \emph {et~al.}(2020)\citenamefont
  {Munteanu}, \citenamefont {Suvorov}, \citenamefont {Dolan}, \citenamefont
  {Sgalaberna}, \citenamefont {Bolognesi}, \citenamefont {Manly}, \citenamefont
  {Yang}, \citenamefont {Giganti}, \citenamefont {Iwamoto},\ and\ \citenamefont
  {Jesús-Valls}}]{Munteanu:2019llq}%
  \BibitemOpen
  \bibfield  {author} {\bibinfo {author} {\bibfnamefont {L.}~\bibnamefont
  {Munteanu}}, \bibinfo {author} {\bibfnamefont {S.}~\bibnamefont {Suvorov}},
  \bibinfo {author} {\bibfnamefont {S.}~\bibnamefont {Dolan}}, \bibinfo
  {author} {\bibfnamefont {D.}~\bibnamefont {Sgalaberna}}, \bibinfo {author}
  {\bibfnamefont {S.}~\bibnamefont {Bolognesi}}, \bibinfo {author}
  {\bibfnamefont {S.}~\bibnamefont {Manly}}, \bibinfo {author} {\bibfnamefont
  {G.}~\bibnamefont {Yang}}, \bibinfo {author} {\bibfnamefont {C.}~\bibnamefont
  {Giganti}}, \bibinfo {author} {\bibfnamefont {K.}~\bibnamefont {Iwamoto}}, \
  and\ \bibinfo {author} {\bibfnamefont {C.}~\bibnamefont {Jesús-Valls}},\
  }\href {\doibase 10.1103/PhysRevD.101.092003} {\bibfield  {journal} {\bibinfo
   {journal} {Phys. Rev. D}\ }\textbf {\bibinfo {volume} {101}},\ \bibinfo
  {pages} {092003} (\bibinfo {year} {2020})},\ \Eprint
  {http://arxiv.org/abs/1912.01511} {arXiv:1912.01511 [physics.ins-det]}
  \BibitemShut {NoStop}%
\bibitem [{\citenamefont {Hamacher-Baumann}\ \emph {et~al.}(2020)\citenamefont
  {Hamacher-Baumann}, \citenamefont {Lu},\ and\ \citenamefont
  {Martín-Albo}}]{Hamacher-Baumann:2020ogq}%
  \BibitemOpen
  \bibfield  {author} {\bibinfo {author} {\bibfnamefont {P.}~\bibnamefont
  {Hamacher-Baumann}}, \bibinfo {author} {\bibfnamefont {X.}~\bibnamefont
  {Lu}}, \ and\ \bibinfo {author} {\bibfnamefont {J.}~\bibnamefont
  {Martín-Albo}},\ }\href {\doibase 10.1103/PhysRevD.102.033005} {\bibfield
  {journal} {\bibinfo  {journal} {Phys. Rev. D}\ }\textbf {\bibinfo {volume}
  {102}},\ \bibinfo {pages} {033005} (\bibinfo {year} {2020})},\ \Eprint
  {http://arxiv.org/abs/2005.05252} {arXiv:2005.05252 [physics.ins-det]}
  \BibitemShut {NoStop}%
\end{thebibliography}%

\clearpage

\end{document}